\documentclass[aps,prd,twocolumn,titlepage,nofootinbib]{revtex4}
\usepackage{graphicx,physics}
\usepackage{color}
\usepackage{dcolumn,comment}
\usepackage{latexsym}

\usepackage[normalem]{ulem}
\usepackage{hyperref,amssymb}
\usepackage{url}
\usepackage{color,soul}
\usepackage{graphicx}
\usepackage{verbatim}
\usepackage{multirow}
\usepackage{amsmath}
\newcommand{\beq}{\begin{eqnarray}}
\newcommand{\eeq}{\end{eqnarray}}
\usepackage{mathrsfs}
\usepackage{float,soul}
\usepackage[dvipsnames]{xcolor}
\usepackage{mathtools}
\usepackage{slashed}
\usepackage{physics}	
\usepackage{graphicx}   
\usepackage{epstopdf}
\usepackage{subfigure}  
\usepackage{bbold}
\usepackage{wasysym}
\usepackage{feynmp}
\usepackage{hyperref}
\usepackage{tabularx}

\hypersetup{colorlinks,
}
\usepackage{bm}

\usepackage{cancel}
\begin{document}

\title{ \large Resonant Coupling and the Non-Phononic Flat Band in Amorphous Solids}
\author{Matteo Baggioli$^{1,2}$}
\email{b.matteo@sjtu.edu.cn}
\author{Bingyu Cui$^{3}$}
\email{bycui@cuhk.edu.cn}
\address{$^1$School of Physics and Astronomy, Shanghai Jiao Tong University, Shanghai 200240, China}
\address{$^2$Wilczek Quantum Center, School of Physics and Astronomy, Shanghai Jiao Tong University, Shanghai 200240, China}
\address{$^3$School of Science and Engineering, The Chinese University of Hong Kong (Shenzhen), Longgang, Shenzhen, Guangdong, 518172, P.R. China}

\begin{abstract}
Recent experiments and simulations provide compelling evidence for the emergence of a non-phononic flat band in the dynamical structure factor of two- and three-dimensional amorphous solids. This feature has been suggested to be connected to the excess in the reduced vibrational density of states of glasses, commonly known as the boson peak, and displays several apparently universal characteristics. First, it is nearly dispersionless, with an energy close to the boson-peak frequency. Second, its intensity is negligible below a critical wave vector of the order of the first diffraction peak. Third, its reduced intensity exhibits a strong correlation with the static structure factor. Here, we revisit the resonant-coupling model, a single-mode harmonic realization of the soft-potential scenario in which acoustic phonons interact with single frequency quasi-localized vibrations. We show that, under minimal phenomenological assumptions, this framework naturally reproduces the main features of the observed flat band and clarifies its connection to the boson peak.
\end{abstract}
\maketitle

\section*{Introduction}
Amorphous solids occur in many forms and across many length scales in nature, ranging from metallic glasses and colloidal suspensions to granular matter and polymers. Their vibrational and thermodynamic properties deviate markedly from those of ideal crystals, whose behavior is well described by textbook theories, e.g. the Debye model, \cite{chaikin1995principles}. This departure gives rise to a variety of glassy anomalies that have been extensively studied over the past decades \cite{ramos2022low,phillips1981amorphous,RevModPhys.74.991,Esquinazi1998}. Prominent examples include the boson-peak (BP) excess in the Debye-reduced heat capacity and vibrational density of states, the linear-in-temperature scaling of the heat capacity in the low-temperature limit, the plateau observed in the thermal conductivity, and the anomalous sound attenuation, among others. 

\begin{figure*}
    \centering
    \includegraphics[width=\linewidth]{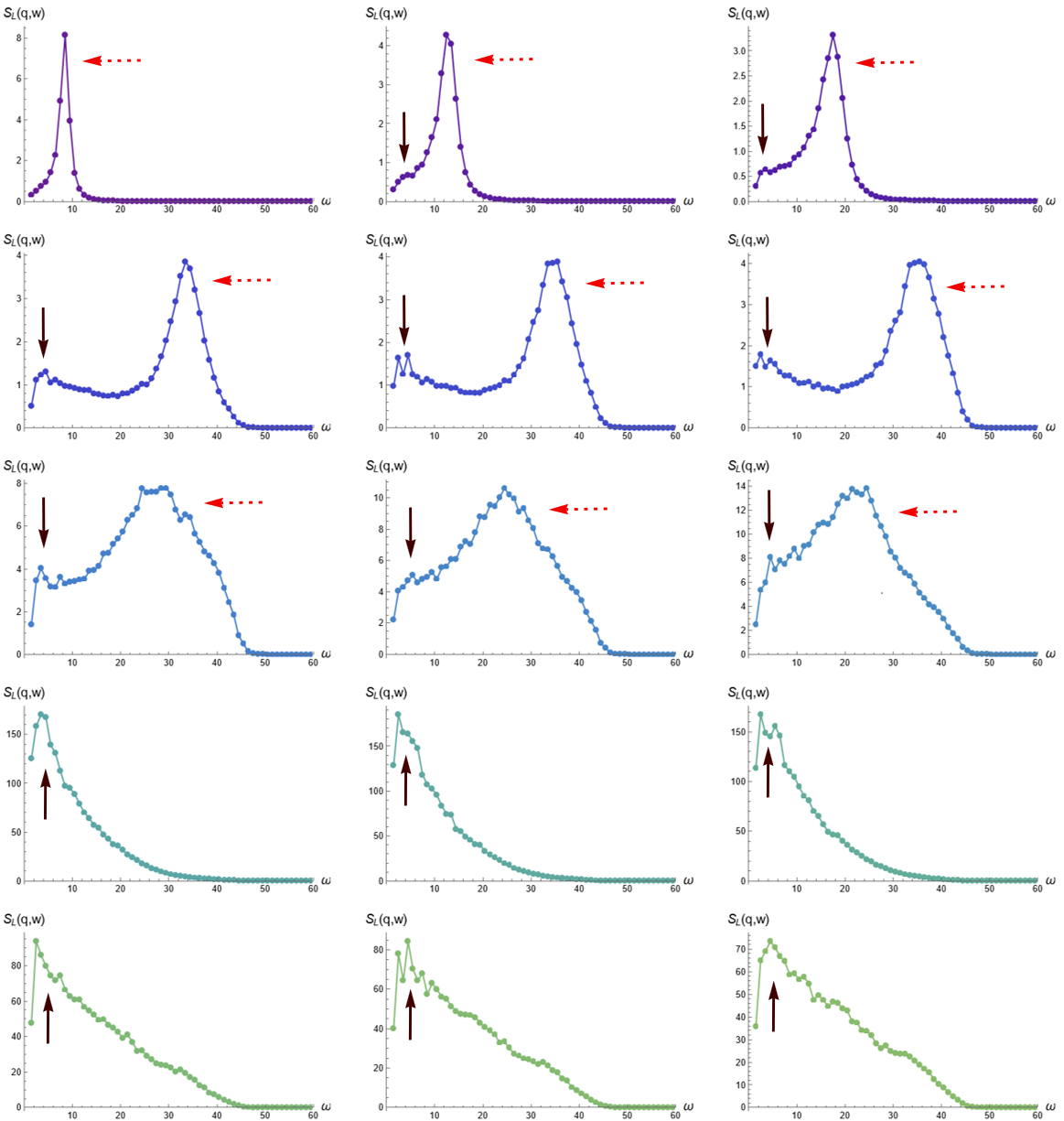}
    \caption{\textbf{The emergence of the non-phononic flat band in the dynamic structure factor of amorphous solids: the example of metallic glasses.} Constant-$q$ cuts of the longitudinal dynamical structure factor $S_L(q,\omega)$ as a function of frequency $\omega$ for a simulated three-dimensional metallic glass. The wave vector $q$ increases from the top-left panel to the bottom-right panel; $q=0.1,0.2,0.3,0.9,1,1.1,1.7,1.8,1.9,2.5,2.6,2.7,3.2,3.3,3.4$. The red dashed arrow indicates the dispersing phononic peak, while the filled black arrow marks the non-phononic peak associated with the flat band. The data are reproduced with permission from Ref.~\cite{Mahajan2025FlatBand}.}
    \label{fig1}
\end{figure*}

Only in recent years (see however \cite{FLUBACHER195953,Tsuneyoshi_Nakayama_2002} for similar older observations), combined experimental and numerical studies \cite{tanakaNatPhys,tanakaPRR,Mahajan2025FlatBand,Li2025,mizuno2025bosonpeakcovalentnetwork,mizuno2026bosonpeakdynamicalstructure,PhysRevLett.133.188302,PhysRevB.77.214309,Ruzicka04,BaldiChapter,Tomterud2023,PhysRevLett.99.035503,PhysRevB.111.205423,PhysRevB.53.12107,inoue1991low,PhysRevLett.96.045502} have revealed the emergence of another anomalous feature that appears to be universal across a wide range of amorphous systems. In particular, it has been observed that the dynamical structure factor $S_L(q,\omega)$ and its transverse counterpart $S_T(q,\omega)$ \cite{doi:10.1142/9781800612587_0010}, in addition to the expected phononic excitations, display a pronounced non-phononic signal. Figure \ref{fig1} reproduces simulation data for a three-dimensional metallic glass from \cite{Mahajan2025FlatBand}, where this feature can be clearly observed as indicated by the vertical black arrow. This contribution, which we will refer to as the \emph{flat band}, exhibits several intriguing and apparently universal characteristics. 

First, its characteristic energy $\omega_{\mathrm{flat}}$ is nearly independent of the wave vector $q$, hence the designation \emph{flat} or dispersionless. Second, this energy scale closely coincides with the boson-peak frequency, defined as the position of the maximum in the Debye-reduced vibrational density of states $D(\omega)/\omega^{d-1}$, where $d$ denotes the number of spatial dimensions. Third, the intensity of the flat-band signal is extremely weak, if not vanishing, below a critical wave vector $q^*$, which is of the same order as the position of the first diffraction peak in the static structure factor $S(q)$. Finally, the reduced intensity of the flat band in the dynamical structure factor exhibits a strong correlation with the static structure factor $S(q)$ itself. 

Each of these properties carries important physical insights, which will be discussed in detail throughout this work. Taken together, these recent observations call for a theoretical framework capable of capturing all these features in a unified way. A natural starting point for such a framework is the resonant coupling model, which is based on the existence and role of quasi-localized vibrations (QLVs), or resonant vibrations, in glasses.

Motivated by early simulations revealing localized soft modes \cite{Buchenau01021992,PhysRevLett.66.636,PhysRevB.44.6746} and experiments pointing to additional non-phononic excitations \cite{PhysRevLett.77.4035,PhysRevLett.60.1318}, the effects of QLVs on the low-temperature properties of glasses, including heat capacity, phonon scattering, and thermal conductivity, have been extensively studied within the soft-potential model (SPM) framework \cite{Karpov1983,Karpov1985,PhysRevB.43.5039,Gurevich03,Parshin07,Buchenau1992}.

Around the same period, Schober and collaborators performed detailed numerical simulations \cite{PhysRevLett.66.636,PhysRevB.53.11469,PhysRevB.44.6746,SCHOBER1993965,PhysRevB.62.3181,Schober_2004} and revisited the SPM by introducing a simplified harmonic description in which an amorphous solid is modeled as an elastic continuum coupled to single-frequency quasi-localized vibrations (QLVs) \cite{Schober2011}. This approach, which we refer to hereafter as the \emph{resonant coupling model} (RCM), provides a theoretical framework in which the boson peak, together with anomalous phonon properties such as enhanced sound attenuation, emerges naturally as a resonance effect arising from the coupling between acoustic phonons and harmonic QLVs.

The resonant-coupling model corresponds to a specific limit of the more general SPM. More precisely, it is obtained within the harmonic approximation by assuming a very narrow distribution of the quadratic potential parameter, resulting in a single quasi-localized vibration with a well-defined frequency.

More recently, this scenario, and the broader idea that several anomalous properties of glasses, in particular the boson peak (BP), may originate from the hybridization between phonons and quasi-localized excitations, has been revisited and developed in various forms. One line of work focuses on quasi-localized modes identified in the vibrational spectrum through harmonic normal mode analysis \cite{10.1063/5.0069477}, supporting the view that the excess modes responsible for the BP and the associated anomalous sound attenuation originate from such excitations and their coupling to acoustic phonons \cite{10.1063/5.0147889,PhysRevResearch.6.023053,10.1063/5.0246261,shivamPRL,ch39-6bhs,shivamSciPost,doi:10.1126/sciadv.adu6097,Flenner2025}. A different perspective attributes the BP to approximately one-dimensional, string-like excitations emerging around the BP frequency and possibly coupling to acoustic phonons \cite{10.1063/5.0039162,10.1063/5.0210057,Jiang_2024,tanakaNatPhys,tanakaPRR,Xi2026}. This latter picture bears strong similarities to the elastic string theory proposed by Lund and collaborators \cite{PhysRevB.101.174311,lund2026quantumtheoryelasticstrings} and previous approaches based on intermediate length scale clusters \cite{PhysRevB.45.2490} (see also \cite{MALINOVSKY199163}). Interestingly, the idea of acoustic phonons coupled to local oscillators has also been compared with the predictions of heterogeneous elasticity theory, leading to closely related results for the vibrational density of states \cite{Maurer2004}.

Despite these developments, most resonant-coupling approaches and related frameworks have focused on explaining the excess in the Debye-reduced vibrational density of states and the associated heat-capacity anomaly known as the boson peak. In contrast, wave-vector--dependent observables, most notably the dynamic structure factor $S(q,\omega)$, have remained largely unexplored. Only in the past few years have theoretical studies begun to address the emergence of a flat-band signal in $S(q,\omega)$, for instance within stringlet-based models \cite{Jiang_2024} and effective-medium mean-field approaches \cite{mizuno2025bosonpeakcovalentnetwork}. A systematic theoretical analysis of this phenomenon, however, is still lacking. 

In this work, we revisit the resonant-coupling model \cite{Schober2011} and reexamine it in the context of recent experimental and numerical observations of a non-phononic flat band in the dynamical structure factor. For a recent review of these observations, we refer the reader to Ref.~\cite{Mahajan2025FlatBand}. 

\section*{The resonant coupling model}
As our starting point, we consider a renormalized phonon Green function whose inverse is given by
\begin{equation}
G_\alpha^{-1}(q,\omega)=\omega^{2}-\Omega_\alpha(q)^{2}+i\,\omega\,\Gamma_\alpha(q)
-\frac{g_\alpha(q)}{\omega^{2}-\omega_{\mathrm{QLV}}^{2}+i\,\omega\,\gamma}
\label{eq:Ginv}
\end{equation}
where $\alpha=L,T$ indicates the phonon polarization, while $\omega,q$ are respectively the frequency and the wavevector. Here, $\Omega_\alpha(q)$ is the bare dispersion that, in the low-wavevector limit, reduces to $\Omega_\alpha(q)=v_\alpha q$. Additionally, $\Gamma_\alpha(q)$ represents a phonon damping term that accounts for scattering mechanisms, such as phonon--phonon interactions or disorder-induced scattering. This term must vanish in the limit $q \rightarrow 0$. The parameter $\omega_{\mathrm{QLV}}$ denotes the characteristic frequency of the QLVs (the \textit{resonance frequency}), while $\gamma$ is the associated damping, which for simplicity is assumed to be independent of both $q$ and $\omega$. Finally, the last term in Eq. \eqref{eq:Ginv} is the standard resonant self-energy generated by coupling an acoustic excitation to a damped quasi-localized oscillator, in which $g_\alpha(q)$ characterizes the coupling between the acoustic phonons and the QLVs that has also to vanish in the $q \rightarrow 0$ limit. 

The model defined by Eq.~\eqref{eq:Ginv} can be derived from first principles along different routes. Schober and collaborators~\cite{Schober2011} obtained it by considering a Hamiltonian $H$ composed of three contributions: a term $H_{\mathrm{ph}}$ describing extended acoustic phonons, a term $H_{\mathrm{lv}}$ accounting for quasi-localized vibrations (QLVs), and an interaction term $H_{\mathrm{int}}$ capturing both the coupling between phonons and QLVs and any residual interactions among the QLVs themselves. Using the standard $T$-matrix formalism for phonon scattering~\cite{maradudin1971theory}, one then derives the phonon Green's function, recovering the form of Eq.~\eqref{eq:Ginv}. Within this framework, QLVs originate from local deviations of the force-constant matrix from its average value in the glass, as well as possible local mass fluctuations~\cite{Buchenau1992,Gurevich2003}.

Recently, Tanaka and collaborators \cite{Xi2026}, derived the same form using a stress-strain coupling framework. More generally, the same structure emerges whenever acoustic phonons are coupled to non-phononic excitations, and has appeared in various contexts, including string-inspired models of the boson peak~\cite{PhysRevB.101.174311,Jiang_2024}.

As a general remark, the microscopic nature of the defects primarily enters through the coupling $g_\alpha(q)$. By contrast, $\Gamma(q)$ acts as an effective phenomenological damping parameter for the phonon modes, encompassing various microscopic mechanisms such as phonon--phonon scattering, disorder-induced broadening, and related dissipative processes.

We emphasize that Eq.~\eqref{eq:Ginv} implies that the phonon self-energy arising from the coupling to the QLVs takes the simple resonant form
\begin{equation}
    \Sigma_\alpha(q,\omega)
    =
    \frac{g_\alpha(q)}
    {\omega^{2}-\omega_{\mathrm{QLV}}^{2}+i\,\omega\,\gamma}.
\end{equation}
The dynamical structure factor can be approximated, in the classical limit, as \cite{FetterWalecka1971}
\begin{equation}\label{dyn}
S_{\alpha}(q,\omega)
=
\frac{2 q^2 k_B T}{M \omega \pi}
\,\mathrm{Im}\!\left[G_\alpha(q,\omega)\right].
\end{equation}
Here, $\alpha=L$ corresponds to the usual dynamical structure factor associated with density--density fluctuations. In contrast, $\alpha=T$ denotes the transverse counterpart. The quantity $M$ denotes the particle mass (or the effective mass associated with the vibrating units of the system).

Finally, Eq.~\eqref{eq:Ginv} assumes a single QLV frequency and damping. To account for multiple QLVs, consistent with recent studies \cite{10.1063/5.0147889,PhysRevResearch.6.023053}, one can naturally introduce a continuous self-energy contribution:
\begin{equation}
  \Sigma_\alpha(q, \omega) = \int_0^{\infty} \frac{\mathcal{C}_\alpha(q, \omega') \, g_{\mathrm{QLV}}(\omega')}{\omega^2 - \omega'^2 + i\omega\gamma(\omega')} \, d\omega'.  
\end{equation}
Here, $g_{\mathrm{QLV}}(\omega')$ is the density of states (DOS) of the QLVs, $\mathcal{C}_\alpha(q, \omega')$ is the momentum- and frequency-dependent phonon–QLV coupling, and $\gamma(\omega')$ represents an intrinsic frequency-dependent damping. In the single-mode limit,
\begin{equation}
    g_{\mathrm{QLV}}(\omega') = \delta(\omega' - \omega_{\mathrm{QLV}}), \qquad \gamma(\omega')=\gamma,
\end{equation}
and $\mathcal{C}_\alpha(q, \omega') = g_\alpha(q)$, this expression directly recovers the single-mode approximation used in this manuscript.

While this continuum formulation provides a more realistic description of disordered systems, it introduces additional functional degrees of freedom in $g_{\mathrm{QLV}}(\omega')$ and $\mathcal{C}_\alpha(q, \omega')$ to be determined. Although these can be partially constrained, for example, by enforcing the low-frequency non-phononic scaling $g_{\mathrm{QLV}}(\omega') \propto \omega'^4$ \cite{10.1063/5.0246261} or via microscopic modeling, applying this more general framework requires further input, which lies beyond the scope of this work.

\section*{Results}
To avoid clutter, we omit the polarization label $\alpha$ in what follows and reintroduce it only when relevant.
\subsubsection*{General remarks}
The dispersion relations of the excitations within the hybrid phonon--QLV model follow from the poles of the Green's function in Eq.~\eqref{eq:Ginv}, i.e., from the solutions of
\begin{equation}
    \left(\omega^{2}-\Omega(q)^{2}+i\,\omega\,\Gamma(q)\right)
    \left(\omega^{2}-\omega_{\mathrm{QLV}}^{2}+i\,\omega\,\gamma\right)
    = g(q).\label{fullfu}
\end{equation}

In the absence of coupling, $g_\alpha(q)=0$, acoustic phonons decouple from the QLVs. Their properties follow from the dispersion relation
\begin{equation}
    \omega=\frac{1}{2}\left(-i \Gamma(q) + \sqrt{4 \Omega^{2}(q)-\Gamma^{2}(q)}\right).
    \label{eq:phonon_dispersion}
\end{equation}
Here $\Omega(q)$ denotes the bare phonon dispersion, while $\Gamma(q)$ is the corresponding linewidth. By symmetry, both quantities vanish in the limit $q \to 0$. In the long-wavelength regime, one recovers the linear dispersion 
$\Omega(q)=v q + \dots$, as expected from viscoelasticity theory~\cite{chaikin1995principles}. At finite temperature, where phonon--phonon scattering is active, one typically finds 
$\Gamma(q)\sim q^2$ in the hydrodynamic regime \cite{chaikin1995principles}, consistent with Akhiezer damping~\cite{akhiezer1939sound}. 
In contrast, at zero temperature the attenuation is dominated by elastic disorder, leading to a Rayleigh regime 
$\Gamma(q)\sim q^4$ at low $q$. This scaling can be understood microscopically in terms of scattering from 
static inhomogeneities, as described by Klemens~\cite{klemens1951scattering}. At finite temperature, the Rayleigh regime may still persist over an intermediate range of wave vectors before crossing over to the $q^2$ behavior at larger $q$ (see, e.g., \cite{D0SM00633E}).

\begin{figure}[ht]
    \centering
    \includegraphics[width=0.8\linewidth]{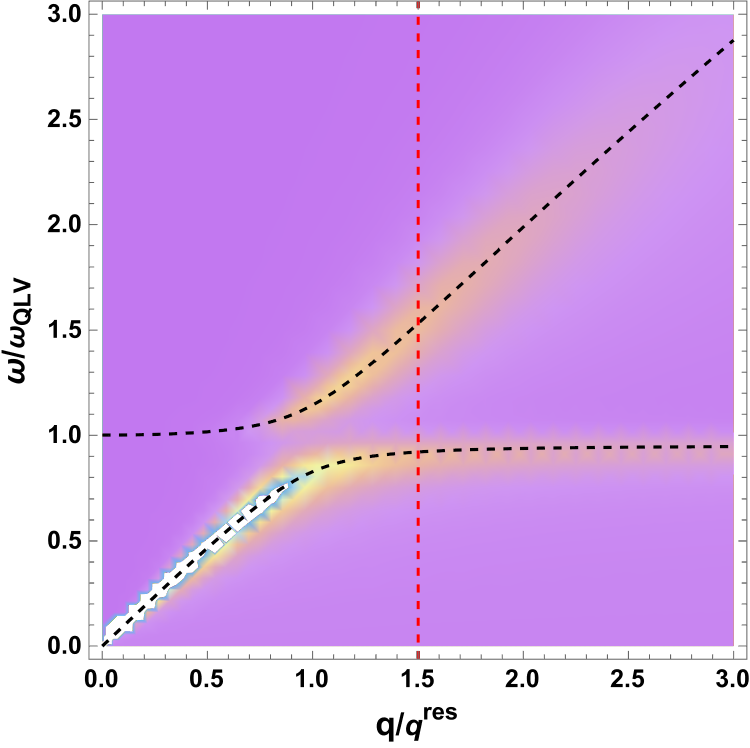}

    \vspace{0.1cm}
    
    \includegraphics[width=0.8\linewidth]{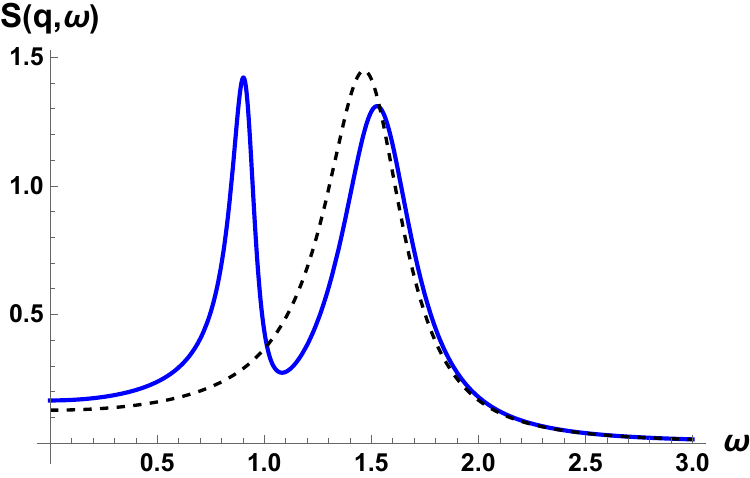}
    \caption{\textbf{Example of the hybridized excitation spectrum.} \textbf{Top:} Density map of the dynamic structure factor $S(q,\omega)$ and dispersion relation of the modes (black dashed curves) showing the avoided crossing and frequency splitting around $q^{\mathrm{res}}=1$. \textbf{Bottom:} dynamic structure factor at fixed $q=1.5$ (red vertical dashed line in top panel), displaying the characteristic two-peak structure, which is absent in the limit of vanishing coupling to QLVs (black dashed curve). Parameters: $\omega_{LV}=k_BT=M=1$ with $\Omega(q)=q$, $\gamma=0.1$, $\Gamma(q)=0.2 q^2$ and $g(q)=0.1 q^2$. The power-law scalings of $\Gamma(q),g(q)$ do not carry any specific physical meaning at this stage.}
    \label{fig2}
\end{figure}

Assuming that the damping coefficients $\Gamma(q)$ and $\gamma$ are negligible compared to $\Omega(q)$ and $\omega_{\mathrm{QLV}}$, the equation for the dispersion relation of the modes, Eq.~\eqref{fullfu}, reduces to
\begin{equation}
(\omega^2 - \Omega^2(q))(\omega^2 - \omega_{\mathrm{QLV}}^2) = g(q).
\label{eq:resont}
\end{equation}
In the absence of coupling, $g(q)=0$, the two modes (acoustic phonons and QLVs) are independent and given by $\omega=\Omega(q)$ and $\omega=\omega_{\mathrm{QLV}}$. For weak coupling, hybridization occurs in the vicinity of the crossing point of the uncoupled modes. This defines the resonance condition
\begin{equation}
\Omega(q^{\mathrm{res}})=\omega_{\mathrm{QLV}}.
\end{equation}
For a linear phonon dispersion at low $q$, $\Omega(q)=v q$, this reduces to
\begin{equation}
q^{\mathrm{res}}=\frac{\omega_{\mathrm{QLV}}}{v}.\label{aa}
\end{equation}
In general, in the limit of weak damping and coupling, the two modes hybridize near resonance, leading to a shift of the frequencies:
\begin{equation}
    \omega_\pm=\pm\left(\omega_{\mathrm{QLV}} + \frac{\sqrt{g(q^{\mathrm{res}})}}{2 \,\omega_{\mathrm{QLV}}}\right).
\end{equation}

An explicit example of the excitation spectrum is shown in Fig.~\ref{fig2}, where a power-law coupling $g(q)\propto q^2$ is assumed and the phonon dispersion is taken to be linear for simplicity. The avoided crossing and the associated frequency splitting around the resonant wave vector $q^{\mathrm{res}}$ are clearly visible in the top panel. In the bottom panel of Fig.~\ref{fig2}, we present a cut at fixed $q$ (dashed red vertical line in the top panel). This reveals the characteristic two-peak structure in the dynamic structure factor, which disappears in the limit of vanishing coupling to the QLVs (black dashed curve in the bottom panel of Fig.~\ref{fig2}). Note that Fig.~\ref{fig2} illustrates the generic hybridization expected when the coupling remains finite at the resonance wave vector. The experimentally relevant regime discussed below instead corresponds to a coupling that is suppressed near \(q^{\mathrm{res}}\).
\subsubsection*{Flat band intensity and constraints on the resonant coupling}
So far, we have not imposed any specific constraints on the model, aside from assuming that the damping terms are negligible. However, simulations and experimental observations can be used to further constrain the theoretical framework introduced above, in particular the form of the resonant coupling $g(q)$. 

From the existing experimental and simulation results (see \cite{Mahajan2025FlatBand} for a review), several robust features have now emerged:

\begin{enumerate}
    \item[(i)] The intensity of the flat-band signal in $S(q,\omega)$ vanishes below a critical wavevector $q^*$, which appears to lie close to the first diffraction peak of the static structure factor $S(q)$~\cite{Li2025};
    
    \item[(ii)] In contrast to the simple scenario illustrated in Fig.~\ref{fig2}, no clear avoided-crossing behavior is observed in the dynamic structure factor;
    
    \item[(iii)] The $q$-dependent intensity of the flat band is strongly correlated with the shape of the static structure factor $S(q)$.
\end{enumerate}

Within the resonant-coupling model, conditions (i) and (ii) imply that
\begin{align}
& g(q) \approx 0 \qquad \text{for } q<q^*, \qquad \text{with } q^*>q^{\mathrm{res}}.
\end{align}
In other words, the resonant coupling between acoustic phonons and QLVs must be strongly suppressed at long wavelengths and become effective only at shorter length scales, set approximately by \(1/q^*\). Under these assumptions, the flat band remains close to \(\omega_{\mathrm{QLV}}\), while acquiring a small shift and broadening due to its hybridization with the phonon. Moreover, no avoided crossing is induced.

\begin{figure}[ht]
    \centering
    \includegraphics[width=0.8\linewidth]{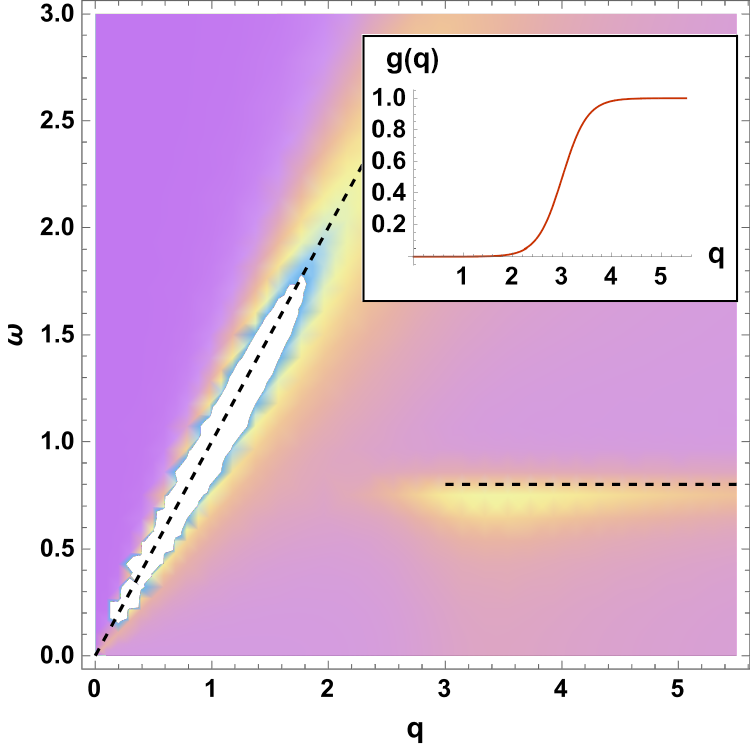}
    \caption{\textbf{Resonant coupling and flat-band intensity.} Example of a resonant coupling $g(q)$ that vanishes below a critical wavevector $q^*\approx 2.5$ (inset). The corresponding density map of $S(q,\omega)$ displays a flat-band signal emerging only for $q>q^*$. Here, $M=k_B T=1$, $\Omega(q)=q$, $\Gamma(q)=0.2q^2$, $\gamma=0.15$, and $\omega_{\mathrm{QLV}}=0.8$ (horizontal black dashed line).}
    \label{fig4}
\end{figure}

An illustrative example that takes into account these two constraints (i)-(ii) is shown in Fig.~\ref{fig4}, where we consider a simple form of the coupling \(g(q)\) that vanishes below \(q^* \approx 2.5\) (in the chosen units). As evident from the heat map of \(S(q,\omega)\), the features described in (i)-(ii) are reproduced. In particular, the flat-band energy remains very close to the resonant frequency \(\omega_{\mathrm{QLV}}\) (horizontal black dashed line), as anticipated, and its intensity vanishes for $q< q^*$. Moreover, no avoided crossing is observed, since by construction the coupling between the two modes vanishes at the resonant wave vector.

What remains to be understood is how to incorporate, within this theoretical framework, the empirical observation that the flat-band intensity closely follows the static structure factor $S(q)$, namely the condition labeled as (iii) in the above list. In this respect, it is important to note that the flat-band intensity depends sensitively on the resonant coupling $g(q)$, and under the conditions discussed below its leading $q$-dependent modulation follows $g(q)$. Therefore, if this correlation is robust and universal, it suggests that the coupling $g(q)$ itself must scale proportionally with $S(q)$.

In general, the intensity of the flat band defined as $S(q,\omega_{\mathrm{QLV}})\equiv f(q)$ is given by:
\begin{equation}
f(q) =
\frac{\xi \,q^2 \left[ g(q) + \gamma \omega_{\mathrm{QLV}}^2 \Gamma(q) \right]}
{\left[
\left( g(q) + \gamma \omega_{\mathrm{QLV}}^2 \Gamma(q) \right)^2
+ \gamma^2 \omega_{\mathrm{QLV}}^2 \left( \omega_{\mathrm{QLV}}^2 - \Omega(q)^2 \right)^2
\right]}.
\end{equation}
with 
\begin{equation}
    \xi=\frac{2 k_B T \gamma }{M \pi}.
\end{equation}
We now discuss the general conditions under which the full expression $f(q)$ is proportional to $g(q)$ up to a power of $q$, without specifying the functional form of $\Omega(q)$ and $\Gamma(q)$. A scaling of the form $f(q)\propto q^m g(q)$ (with $m=2$ in the present case due to the explicit prefactor) is obtained provided two main conditions are satisfied. First, the numerator must be dominated by $g(q)$, which requires
\begin{equation}\label{cond1}
g(q)\gg \gamma \omega_{\mathrm{QLV}}^2 \Gamma(q).
\end{equation}
Second, the denominator must be effectively independent of the detailed $q$-dependence of $g(q)$, which is achieved when the term proportional to $(\omega_{QLV}^2-\Omega(q)^2)^2$ dominates. This requires
\begin{equation}\label{cond2}
    \left(g(q)+\gamma \omega_{\mathrm{QLV}}^2 \Gamma(q)\right)^2 \ll \gamma^2 \omega_{\mathrm{QLV}}^2 \left(\omega_{\mathrm{QLV}}^2-\Omega(q)^2\right)^2.
\end{equation}
In addition, one must remain away from the resonance condition $\Omega(q)\approx \omega_{\mathrm{QLV}}$, where the denominator develops strong $q$-dependence and the scaling breaks down. Under these conditions, the denominator is controlled by the detuning term. If this detuning varies slowly over the $q$-range of interest compared with the oscillatory structure of $g(q)$, then the dominant modulation of $f(q)/q^2$ follows $g(q)$, namely
\begin{equation}
f(q)\propto q^2 g(q).\label{ee}
\end{equation} Deviations from this behavior arise when the dissipative contribution $\Gamma(q)$ dominates the numerator, when the $(g(q)+\cdots)^2$ term controls the denominator, or in the vicinity of the resonance, where the response is no longer governed by a simple scaling form.

Physically, the conditions in Eqs.~(\ref{cond1}-\ref{cond2}) correspond to the regime where the flat band is visible at wave vectors larger than the onset wave vector $q^*$. In this regime, no avoided crossing occurs, and the flat-band frequency remains pinned close to $\omega_{\text{QLV}}$.

Moreover, in the regime where Eq.~\eqref{ee} applies, assuming $g(q)\sim S(q)$, implies that the flat-band intensity $f(q)$ strongly correlates with the static structure factor, consistent with simulation and experimental results \cite{Li2025,Mahajan2025FlatBand}. We note that, since $S(0)\neq 0$, the proportionality discussed above must include an overall factor that vanishes in the $q\to0$ limit in order to ensure that $g(0)\to0$. A natural candidate is a power-law prefactor of the form $\sim q^n$.

\begin{figure}[ht]
    \centering
    \includegraphics[width=0.9\linewidth]{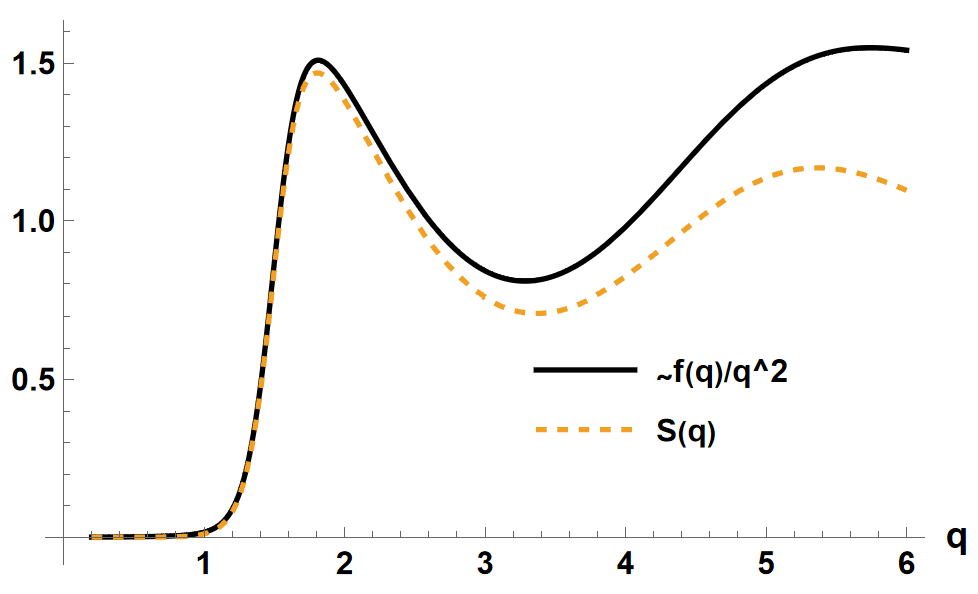}
    \caption{\textbf{Static structure factor and flat-band intensity.} 
A representative example in which the normalized flat-band intensity, $f(q)/q^2$, closely follows the trend of an artificial static structure factor $S(q)$ (orange dashed line), introduced in the model through the simplified choice $g(q)\propto S(q)$. The remaining parameters are set to $k_B T = M = 1$, $\gamma = 0.1$, $\Omega(q) = 0.5\,q$, $\Gamma(q) = 5 \times 10^{-4}\,q^2$, and $\omega_{\mathrm{QLV}} = 10$. 
For clarity of visualization, the black curve is rescaled by an appropriate constant.}
    \label{fig5}
\end{figure}

In Fig.~\ref{fig5}, we present an example in which these constraints are satisfied, demonstrating that the intensity of the flat band closely follows the trend of the static structure factor, as encoded in the coupling \( g(q) \propto S(q) \). 
This analysis shows that the model naturally captures the observed strong correlation between the flat-band intensity and the static structure factor \( S(q) \) \cite{Li2025,Mahajan2025FlatBand}, provided that the coupling \( g(q) \) is taken to be proportional to \( S(q) \). 
The physical interpretation of this choice warrants further investigation and will be briefly discussed in the following sections.

It is worth noting that, within this model, the correlation with the static structure factor is not limited to the intensity of the flat band, but also extends to its frequency. In particular, the position of the flat band, identified as the non-phononic peak in the dynamic structure factor, is not strictly \(q\)-independent and equal to \(\omega_{\mathrm{QLV}}\). Instead, it exhibits a clear \(q\)-modulation, primarily governed by the form of the resonant coupling \(g(q)\), which, as discussed above, is proportional to \(S(q)\).

\begin{figure}
    \centering
    \includegraphics[width=0.7\linewidth]{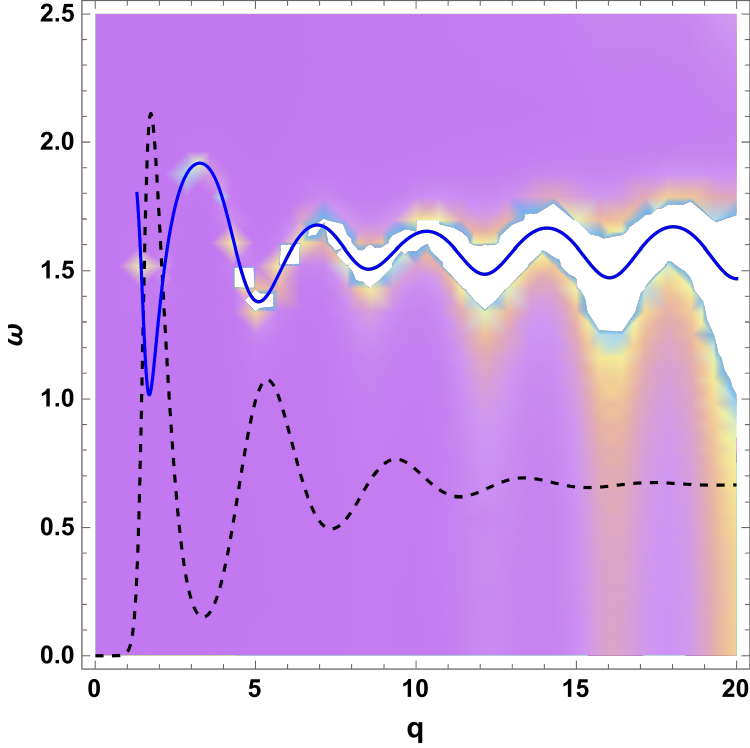}

\vspace{0.2cm}

   \includegraphics[width=0.7\linewidth]{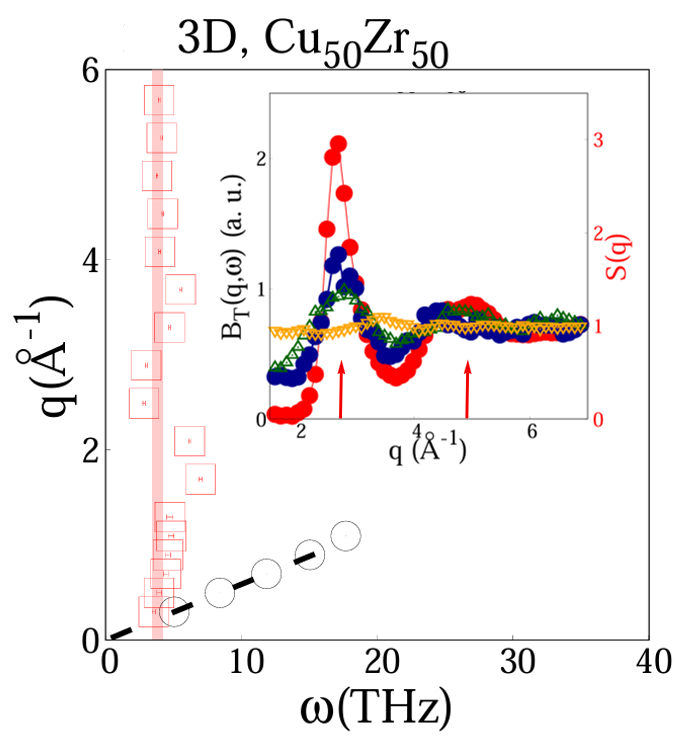}
    \caption{\textbf{Anti-correlation between flat-band dispersion and resonant coupling/static structure factor.}
\textbf{Top panel:} Illustration of the anti-correlation between the flat-band dispersion (color background and solid blue line) and the resonant coupling \(g(q)\) (black dashed line), modeled here as an oscillatory function to mimic the behavior of a realistic static structure factor \(S(q)\). For clarity, the amplitude of \(g(q)\) is rescaled by a factor of 30. The others parameters used in this figure are: $\Gamma=10^{-3}q^2$, $k_B=T=M=1$, $\gamma=0.01$, $\omega_{\mathrm{QLV}}=2$, while the long-wavelength phonon dispersion is taken as $\Omega(q)=3 q +\dots$ The phonon dispersion is not visible because of the small frequency range chose and the higher intensity of the flat band (white signal). \textbf{Bottom panel:} Simulation data from Ref.~\cite{Mahajan2025FlatBand} for a CuZr 3D metallic glass, showing the same anti-correlation between the flat-band dispersion (empty red squares) and \(S(q)\) (red symbols in the inset).}
    \label{fig:new}
\end{figure}

In the top panel of Fig.~\ref{fig:new}, we present an illustrative example in which this feature is intentionally enhanced by isolating the flat-band signal and shifting the phonon dispersion to higher frequencies. There, we can observe a pronounced anti-correlation between \(g(q) \propto S(q)\) (black dashed line) and the flat-band frequency (white signal or blue filled line). The anti-correlation arises from the pole shift near $\omega = \omega_{\text{QLV}}$. Expanding Eq.~\eqref{eq:resont} around $\omega = \omega_{\text{QLV}}$ yields
\begin{equation}
\Delta\omega(q) \simeq \frac{g(q)}{2\,\omega_{\text{QLV}}\,\left(\omega^2_{\text{QLV}} - \Omega(q)^2\right)}.
\end{equation}
Consequently, for $\Omega(q) > \omega_{\text{QLV}}$ (that is the natural hierarchy in the regime where the flat band intensity is non-vanishing), a larger coupling $g(q)$ pushes the flat band downward, driving the anti-correlation with $S(q)$.
When the resonant coupling is large, the flat-band frequency is reduced, following the oscillatory behavior of \(S(q)\). In contrast, at large wave vectors, where \(S(q)\), and thus \(g(q)\), lose their oscillatory character, the flat-band frequency becomes nearly \(q\)-independent and approaches \(\omega_{QLV}\). This is consistent with the fact that, in the large-$q$ limit, the static structure factor approaches unity while its oscillations become progressively damped. As a result, the oscillatory features observed in both the intensity and the frequency of the flat band gradually average out at large $q$.

Interestingly, the trend of the flat-band frequency derived from the resonant-coupling model is consistent with simulation results (see bottom panel of Fig.~\ref{fig:new}) and provides a potential direct test of the theoretical scenario proposed here. We also clarify that Fig.~\ref{fig:new} focuses primarily on the frequency modulation of the flat-band signal. A more in-depth discussion of its intensity, and the connection to experimental and numerical observations, is provided in relation to Fig.~\ref{fig5} above.

\subsubsection*{Two-peak structure in $S(q,\omega)$.}
After discussing in detail several key features of the model, and showing its ability to reproduce the main features observed in experiments and simulations, we now turn to a concrete example illustrating the evolution of the two-peak structure as the wavevector $q$ increases. 
While this behavior can be tracked analytically due to the simplicity of the model, for clarity of presentation we instead focus on an explicit example in which the trend closely resembles the simulation results shown in Fig.~\ref{fig1}.

\begin{figure*}[ht]
    \centering
    \includegraphics[width=\linewidth]{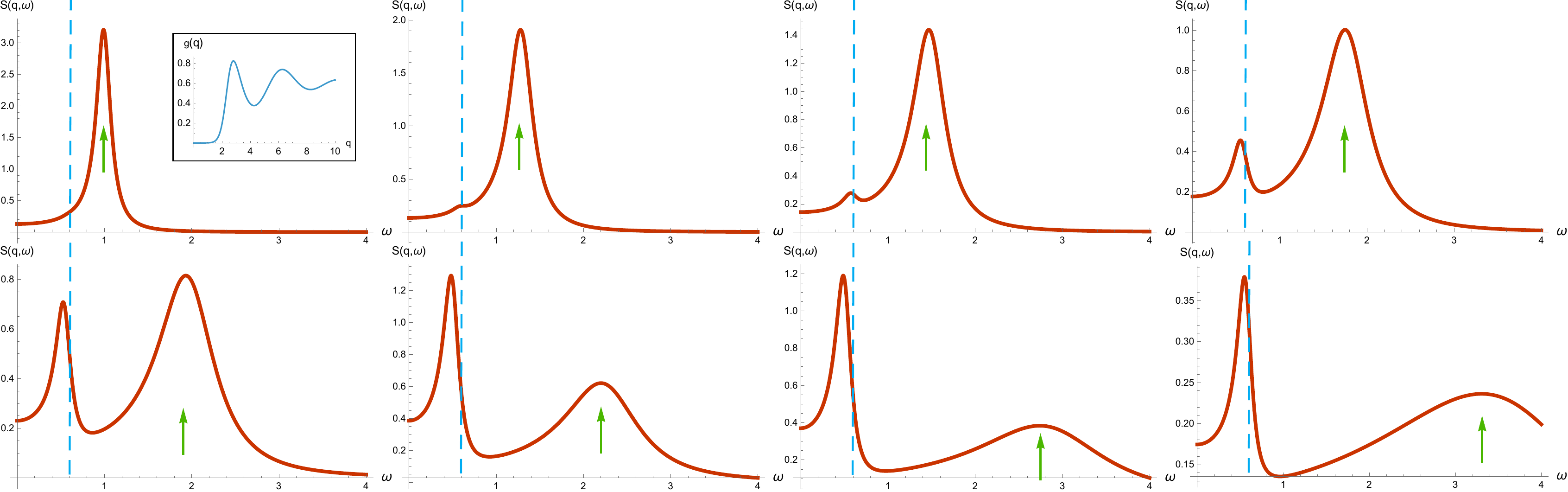}
    \caption{\textbf{The two-peak structure emerging in the theoretical model.} 
A representative example of the evolution of the dynamical structure factor $S(q,\omega)$ as a function of $\omega$ upon increasing $q$ (from the top-left panel to the bottom-right one). 
In this example, the coupling $g(q)$ (see inset in top-left panel) is chosen to be approximately vanishing below a cutoff wavevector $q^* \approx 2.4$. 
Importantly, the frequency of the QLVs is fixed at $\omega_{\mathrm{QLV}} = 0.6$ (vertical blue dashed line) and sets the position of the dispersionless non-phononic signal. 
By contrast, the phonon peak (indicated with a green arrow) disperses with $q$ and broadens due to its increasing damping. The other parameters used in this figure are: $M=k_B T=1$, $\Omega(q)=q$, $\Gamma(q)=0.2 q^2$, $\gamma=0.2$, $\omega_{QLV}=0.6$.}
    \label{fig6}
\end{figure*}

The results are presented in Fig.~\ref{fig6}. At low $q$, the response is dominated by the acoustic phonon peak, while the non-phononic contribution is suppressed, as the coupling $g(q)$ is approximately vanishing for $q \lesssim 2$ in this example (see inset in top-left panel). 
As the wavevector $q$ increases, the intensity of the non-phononic flat-band signal grows, while that of the acoustic phonon decreases. At the same time, the acoustic phonon peak broadens, since its damping increases with $q$. 
For sufficiently large $q$, the non-phononic contribution becomes dominant, and its position remains approximately independent of $q$, i.e., it is flat. Upon further increasing the wavevector (not shown in Fig.~\ref{fig6}), the phononic contribution becomes essentially negligible, leaving a pronounced flat-band signal at approximately $\omega \approx \omega_{\mathrm{QLV}}$ whose intensity is mainly governed by the behavior of the resonant coupling $g(q)$. Overall, these features are in good qualitative agreement with the simulation results shown in Fig.~\ref{fig1}, as well as with other observations reported in the literature discussed above.

\subsubsection*{Boson peak and resonant frequency}
After analyzing the main features of the dynamical structure factor $S(q,\omega)$ and the emergence of the non-phononic flat-band signal, we now turn to the discussion of the corresponding vibrational density of states (VDOS).

The VDOS $D(\omega)$ can be directly derived from the Green's function in Eq.~\eqref{eq:Ginv},
\begin{equation}
D(\omega)= -\frac{2\omega}{\pi}\sum_\alpha \int \frac{d^d q}{(2\pi)^d}\,\Im G_\alpha(q,\omega) .\label{dosdef}
\end{equation}
where $d$ is the spatial dimension of the system.

For simplicity, in the rest of this discussion we will drop the polarization label $\alpha$ and consider a single branch.

\begin{figure*}[ht]
    \centering
    \includegraphics[height=6cm]{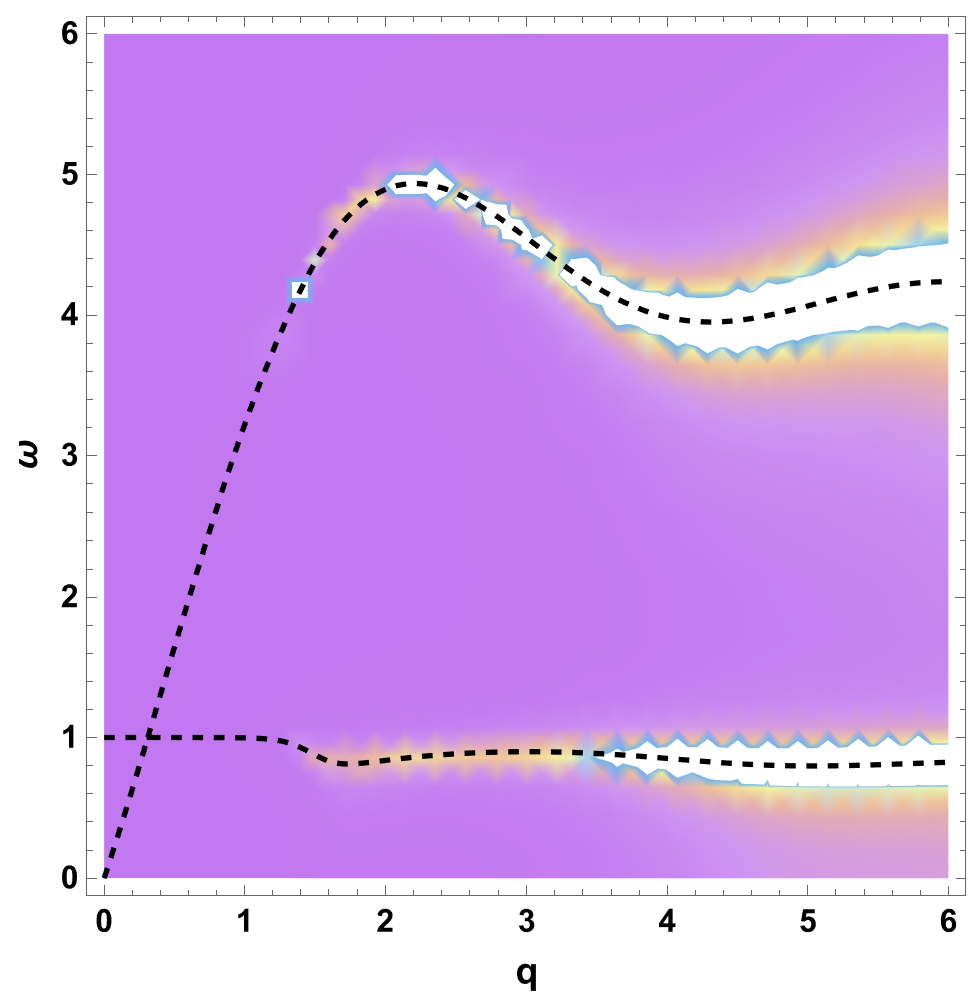}\qquad
 \includegraphics[height=6cm]{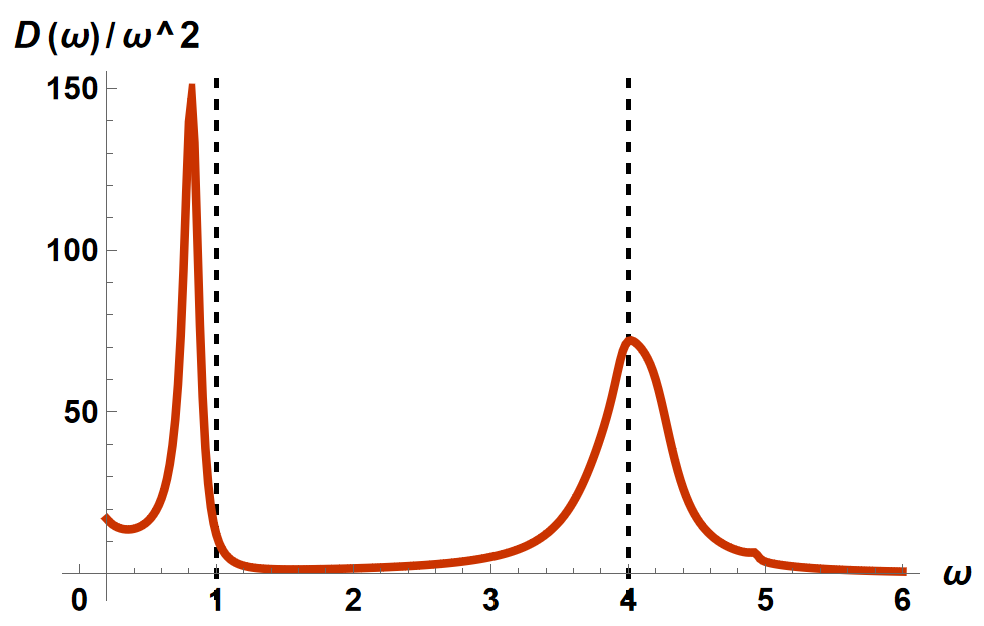}
    \caption{\textbf{Boson peak from a flat band: a concrete example.}
\textbf{Left panel:} Color map of the dynamical structure factor $S(q,\omega)$, showing a dispersing acoustic phonon branch and an approximately flat band, whose spectral weight becomes visible for $q \gtrsim 2$. The bare flat-band frequency is $\omega_{\mathrm{QLV}} = 1$, while the Van Hove frequency of the acoustic branch is $\omega_{\mathrm{VH}} = 4$.  \textbf{Right panel:} The corresponding normalized density of states $D(\omega)/\omega^2$, displaying a boson-peak anomaly near $\omega_{\mathrm{QLV}}$ and a higher-frequency Van Hove peak at $\omega_{\mathrm{VH}}$. The other parameters used for these figures are: $\Gamma(q)=0.01 q^2$, $M=k_B T=1$ and $\gamma=0.1$. In the right panel, the integration range in Eq.~\eqref{dosdef} extends up to a maximum wave-vector $q_{\mathrm{max}}=8$.}
    \label{fig7}
\end{figure*}

In Fig.~\ref{fig7}, we present a benchmark example illustrating the emergence of a boson-peak (BP) anomaly in the three-dimensional Debye-reduced density of states, $D(\omega)/\omega^2$. The BP position matches closely the bare frequency of the flat band in the model, $\omega_{\mathrm{QLV}}$, with a slight downward shift due to damping-induced softening and hybridization effects. This observation imposes a further constraint on the RCM, that is $\omega_{\mathrm{QLV}} \approx \omega_{\mathrm{BP}}$. For clarity, we stress that, within our framework, $\omega_{\text{QLV}}$ is an empirical input chosen near the boson-peak frequency to match the observed alignment with the flat-band energy, rather than a parameter derived from first principles.

We notice that, within the approximation used to define the resonant wave vector in Eq.~\eqref{aa}, the identification $\omega_{\mathrm{QLV}}=\omega_{\mathrm{BP}}$ implies $q^{\mathrm{res}}=\omega_{\mathrm{BP}}/v$. Using simulation data for several amorphous systems reported in Ref.~\cite{Mahajan2025FlatBand} (see bottom panel in Fig.~\ref{fig:new} for the case of the 3D CuZr metallic glass), we verified that $\omega_{\mathrm{BP}}/v$ is systematically smaller than the position of the first diffraction peak, $q^*$, in the static structure factor $S(q)$. Within the theoretical framework developed here, this observation is consistent with the absence of an avoided crossing in the spectrum and therefore provides additional support for the validity of our interpretation.

Importantly, this minimal model makes it clear that the boson peak is not simply the result of a broadened Van Hove singularity, as sometimes suggested (see, e.g., \cite{PhysRevLett.106.225501}). This is consistent with simulations and experimental results proving the independence of these two features, e.g., \cite{PhysRevB.98.174207}. Whether a well-defined Van Hove feature coexists with a genuine boson peak depends primarily on the strength of phonon damping and on the separation of scales between $\omega_{\mathrm{QLV}}$ and $\omega_{\mathrm{VH}}$. In particular, sufficiently strong damping $\Gamma(q)$ broadens the Van Hove singularity to the point that it is no longer distinguishable. For example, in the case considered in Fig.~\ref{fig7}, the phonon damping is chosen to be very small, such that the phonon dispersion retains a large intensity up to very large values of $q$, thereby producing a strong and well-defined Van Hove peak in the corresponding VDOS.

Within this resonant-coupling framework the boson-peak frequency is then primarily set by the flat-band frequency in the dynamical structure factor. This behavior is consistent with observations from both simulations and experiments \cite{tanakaNatPhys,mizuno2026bosonpeakdynamicalstructure,mizuno2025bosonpeakcovalentnetwork,Mahajan2025FlatBand}, and indicates that the flat band plays a central role in the origin of the boson peak in amorphous solids. More broadly, these results support the view that the boson peak is largely of non-phononic origin, emerging from the resonant coupling between non-phononic mode, modeled here as quasi-localized vibrations (QLVs), and acoustic phonons.

In general, we expect the boson peak to also receive contributions from the phononic sector. Whether these contributions are always subleading compared to those arising from the flat band remains an open question. Within our model, the answer depends sensitively on several factors, including the $q$-range over which the flat band exists and its damping. A more direct comparison with numerical simulation models is therefore needed to clarify this issue. Nevertheless, we mention that recent simulation results \cite{mizuno2026bosonpeakdynamicalstructure} support the idea that the boson-peak signal is dominated by the flat-band contribution.

\section*{Discussion and outlook}
In this work, we revisit the resonant-coupling model \cite{Schober2011}, a harmonic single-mode reduction of the more general soft-potential model, in light of recent experimental and numerical observations reporting a universal non-phononic flat-band signal in the dynamic structure factor of two- and three-dimensional amorphous solids (see Ref.~\cite{Mahajan2025FlatBand} for a review).

We have shown that this minimal theoretical framework is able to qualitatively reproduce the key features of this flat band, including its connection to the boson peak. While other approaches, such as marginal-stability frameworks \cite{mizuno2025bosonpeakcovalentnetwork}, may also account for some of these observations, we are currently not aware of any alternative model that captures this phenomenology within such a simple and unified description. In particular, models based exclusively on phononic degrees of freedom are intrinsically unable to account for the emergence of the non-phononic flat band and therefore appear to miss an essential ingredient in the microscopic origin of the boson peak, capturing therefore only its phononic component.

Despite its appealing phenomenology, the resonant-coupling model remains fundamentally macroscopic and phenomenological. In particular, the microscopic origin of the quasi-localized vibrations (QLVs) is not specified, the momentum dependence of the coupling function \(g(q)\) is introduced \emph{ad hoc} rather than derived microscopically, and the characteristic frequency \(\omega_{\mathrm{QLV}}\) enters as a phenomenological scale, typically identified with the boson-peak frequency. These open issues point to the need for a more microscopic understanding of the non-phononic excitations responsible for the flat-band signal.

The goal of the present work is not to derive the microscopic origin of these excitations, but rather to demonstrate that their resonant hybridization with acoustic phonons is sufficient to reproduce the main phenomenology observed experimentally and numerically. More specifically, if amorphous solids host quasi-localized or non-phononic modes in the boson-peak frequency range, and if these modes couple resonantly to acoustic phonons at medium-range-order wave vectors, then a flat-band signal naturally emerges in the dynamical response. Since the full soft-potential model generalizes the simplified resonant-coupling framework considered here, we expect it to capture similar flat-band physics as well, potentially including important anharmonic effects. A detailed investigation along these lines is left for future work.

In this direction, it is worth noting that independent experiments on metallic glasses \cite{tian2021structural} and phase-change materials \cite{qi2026directobservationultrafastamorphousamorphous} have reported real-time observations of collective atomic oscillations associated with medium-range order in amorphous solids, with a coherent frequency matching that of the boson peak. These findings, together with the observed correlation between the flat-band intensity and the static structure factor, point toward a structural origin rooted in medium-range order for both the flat-band signal and, ultimately, the boson peak in amorphous solids. It would be interesting to investigate whether this phenomenology can be understood in terms of acoustic phonons propagating in a disordered environment, as assumed by one of the most successful theories of the boson peak -- the heterogeneous elasticity framework \cite{doi:10.1142/9781800612587_0009}.

Importantly, a central feature of the present model is the coupling $g(q) \propto S(q)$ between acoustic phonons and nonphononic excitations, linking phonon dynamics directly to medium-range structural order. Lacking a first-principles derivation, this relation currently serves as an empirical ansatz required to reproduce simulation and experimental observations. To motivate this empirical scaling, we speculate that $g(q) \propto S(q)$ can be grounded in the framework proposed by Schober, which underpins early formulations of the soft potential model. In this picture, soft nonphononic vibrations stem from local deviations of the force-constant matrix relative to the average glass matrix. Because these fluctuating quantities are anchored to discrete atomic positions, their intensity, and thus their coupling $g(q)$ to acoustic phonons, is naturally mediated by local density fluctuations encoded in Fourier space by $S(q)$. Indeed, early work by Schober and Laird \cite{PhysRevB.44.6746} already suggested the possibility that localized soft modes originate from local structural variations, a connection later elaborated by Schober \cite{Schober_2004}. This hypothesis also aligns with recent findings \cite{li2023spatial}, where nonphononic modes were linked to local shear strain fluctuations whose spatial correlations are dictated by $S(q)$ around the First Sharp Diffraction Peak ($q_{\text{FSDP}}$), reflecting medium-range length scales $\xi_{\text{MRO}} \approx 2\pi/q_{\text{FSDP}}$. While these physical arguments provide a plausible rationale for $g(q) \propto S(q)$, this mechanism remains a theoretical conjecture that requires explicit verification through microscopic modeling or detailed simulations.

Finally, a more detailed analysis of the anomalous phonon attenuation within this theoretical framework would be highly valuable. In this context, it is worth noting that, when projected onto the phonon dispersion \(\omega \propto q\), the attenuation induced by the resonant coupling to QLVs acquires a resonant form, closely resembling that recently used in \cite{Ding2025} to explain the origin of the boson peak in several experimental systems. 

\section*{Acknowledgments}
We thank Shivam Mahajan, Long-Zhou Huang, Cunyuan Jiang, Yun-Jiang Wang, Massimo Pica Ciamarra, Jie Zhang, Peng Tan, Xun-Li Wang, Miguel Angel Ramos, Josep Lluis Tamarit, Huaping Zhang, Yuanchao Hu, Giacomo Baldi, Jack Douglas, Wensi Sun and Hua Tong for useful discussions about the flat band and the BP in amorphous solids.
MB acknowledges the support of the Foreign Young Scholars Research Fund Project (Grant No.22Z033100604) and the sponsorship from the Yangyang Development Fund. B.C. acknowledges the financial support of the National Natural Science Foundation of China (No. 12404232), start-up funding from the Chinese University of Hong Kong, Shenzhen (No. UDF01003468) and the Shenzhen city ``Pengcheng Peacock" Talent Program.

\end{document}